\documentclass[fleqn,usenatbib]{mnras}
\usepackage{newtxtext,newtxmath}
\usepackage[T1]{fontenc}
\usepackage{ae,aecompl}
\usepackage{graphicx}	% Including figure files
\usepackage{amsmath}	% Advanced maths commands
\usepackage{amssymb}	% Extra maths symbols
\usepackage{xspace}
\newcommand{\ngc}{NGC\,6791\xspace}

\title[The blue straggler V106]{The blue straggler V106 in NGC\,6791: \\A prototype progenitor of old single giants masquerading as young.}
\author[K. Brogaard et al.]{K. Brogaard,$^{1,2}$\thanks{E-mail: kfb@phys.au.dk}
S. M. Christiansen,$^{1}$
F. Grundahl,$^{1}$
A. Miglio,$^{2,1}$
R. G. Izzard,$^{3}$
\newauthor 
T. M. Tauris,$^{4,5,6}$
E. L. Sandquist,$^{7}$ 
D. A. VandenBerg,$^{8}$ 
J. Jessen-Hansen,$^{1}$
T. Arentoft,$^{1}$
\newauthor 
H. Bruntt,$^{1}$
S. Frandsen,$^{1}$
J. A. Orosz,$^{7}$ 
G. A. Feiden,$^{9}$ 
R. Mathieu,$^{10}$ 
A. Geller,$^{10}$ 
\newauthor 
M. Shetrone,$^{11}$ 
N. Ryde,$^{12}$ 
D. Stello,$^{1,13,14}$ 
I. Platais,$^{15}$ 
and S. Meibom$^{16}$ 
\\
$^{1}$Stellar Astrophysics Centre, Department of Physics and Astronomy, Aarhus University, Ny Munkegade 120, 8000 Aarhus C, Denmark\\
$^{2}$School of Physics and Astronomy, University of Birmingham, Edgbaston, Birmingham B15 2TT, UK\\
$^{3}$Astrophysics Research Group, Faculty of Engineering and Physical Sciences, University of Surrey, Guildford, GU2 7XH, United Kingdom.\\
$^{4}$Argelander-Institut f\"ur Astronomie, Universit\"at Bonn, Auf dem H\"ugel 71, D-53121 Bonn, Germany\\
$^{5}$Max-Planck-Institut f\"ur Radioastronomie, Auf dem H\"ugel 69, D-53121 Bonn, Germany\\
$^{6}$Department of Physics and Astronomy, Aarhus University, Ny Munkegade 120, 8000 Aarhus C, Denmark\\
$^{7}$Department of Astronomy, San Diego State University, San Diego, CA 92182, USA\\
$^{8}$Department of Physics and Astronomy, University of Victoria, P.O. Box 1700 STN CSC, Victoria, B.C., V8W 2Y2, Canada\\
$^{9}$Department of Physics and Astronomy, Uppsala University, Box 516, 751 20, Uppsala, Sweden\\
$^{10}$Department of Astronomy, University of Wisconsin-Madison, Madison, WI 53706, USA\\
$^{11}$University of Texas, McDonald Observatory, HC75 Box 1337-L Fort Davis, TX 79734, USA\\
$^{12}$Department of Astronomy and Theoretical Physics, Lund Observatory, Lund University, Box 43, 221 00, Lund, Sweden\\
$^{13}$Sydney Institute for Astronomy (SIfA), School of Physics, University of Sydney, NSW 2006, Australia\\
$^{14}$School of Physics, The University of New South Wales, Sydney NSW 2052, Australia\\
$^{15}$Department of Physics and Astronomy, Johns Hopkins University, 3400 North Charles Street, Baltimore, MD 21218, USA\\
$^{16}$Harvard-Smithsonian Center for Astrophysics, Cambridge, MA 02138, USA\\
}

\date{Accepted XXX. Received YYY; in original form ZZZ}

\pubyear{2018}

\begin{document}
\label{firstpage}
\pagerange{\pageref{firstpage}--\pageref{lastpage}}
\maketitle

\begin{abstract}
%This is a simple template for authors to write new MNRAS papers.
%The abstract should briefly describe the aims, methods, and main results of the paper.
%It should be a single paragraph not more than 250 words (200 words for Letters).
%No references should appear in the abstract.

We determine the properties of the binary star V106 in the old open cluster \ngc. We identify the system to be a blue straggler cluster member by using
a combination of ground-based and \textit{Kepler} photometry and multi-epoch spectroscopy. The properties of the primary component are found to be $M_{\rm p}\sim1.67 \rm M_{\odot}$, more massive than the cluster turn-off, with $R_{\rm p}\sim1.91 \rm R_{\odot}$ and $T_{\rm eff}=7110\pm100$ K. The secondary component is highly oversized and overluminous for its low mass with $M_{\rm s}\sim0.182 \rm M_{\odot}$, $R_{\rm s}\sim0.864 \rm R_{\odot}$ and $T_{\rm eff}=6875\pm200$ K. We identify this secondary star as a bloated (proto) extremely low-mass helium white dwarf.
These properties of V106 suggest that it  represents a typical Algol-paradox system and that it evolved through a mass-transfer phase which provides insight into its past evolution. We present a detailed binary stellar evolution model for the formation of V106 using the MESA code and find that the mass-transfer phase only ceased about $40\,{\rm Myr}$ ago. Due to the short orbital period (P=1.4463 d) another mass-transfer phase is unavoidable once the current primary star evolves towards the red giant phase.
We argue that V106 will evolve through a common-envelope phase within the next $100\,{\rm Myr}$ and merge to become a single over-massive giant. The high mass will make it appear young for its true age, which is revealed by the cluster { properties}. Therefore, V106 is potentially a prototype progenitor of old field giants masquerading as young.

% {\color{magenta}or a normal horizontal branch star.}
%We argue that V106 will evolve through a common-envelope phase within the next $100\,{\rm Myr}$ and merge to become a single overmassive giant or one of the extreme horizontal branch stars that \ngc is known to host. 
%If systems like V106 are the typical progenitors of the EHB stars in \ngc this might explain the apparent inconsistency between asteroseismic gravity mode period spacings of the EHB stars and the red clump stars of the cluster. 

\end{abstract}


%%%%%%%%%%%%%%%%%%%%%%%%%%%%%%%%%%%%%%%%%%%%%%%%%%

%%%%%%%%%%%%%%%%% BODY OF PAPER %%%%%%%%%%%%%%%%%%
%TT: Overleaf cashes if there is a page break in the middle of a reference. This can be solved by adding more/less text to avoid this or by a "clearpage" command:
%\clearpage
% * <kfb@phys.au.dk> 2018-04-17T09:44:41.956Z:
%
% ^.
% * <kfb@phys.au.dk> 2018-04-17T09:44:40.170Z:
%
% ^.
\section{Introduction}

%This is a simple template for authors to write new MNRAS papers.
%See \texttt{mnras\_{\rm s}ample.tex} for a more complex example, and \texttt{mnras\_guide.tex}
%for a full user guide.

%All papers should start with an Introduction section, which sets the work
%in context, cites relevant earlier studies in the field by \citet{Others2013},
%and describes the problem the authors aim to solve \citep[e.g.][]{Author2012}.

%Binary star evolution has recieved renewed attention because it has been shown that

A non-neglible fraction of stars in the Milky Way likely evolved through mass-transfer events in binary systems, both in star clusters \citep{Brogaard2016, Handberg2017} and in the field \citep{Jofre2016, Yong2016, Izzard2018}. Thus, there is a need to understand the evolution of such systems to correctly interpret ensemble studies of stars in the Milky Way (MW). 
The study of binary stars allows the determination of the component properties that are often unavailable for single stars. When the binary star is a member of a star cluster, this can be exploited to obtain even more information about the system and constrain uncertain physics in the binary evolution models.
In the following sections we study the binary system V106 in the old open cluster \ngc. We establish cluster membership and the blue straggler star (BSS) nature of the system, and determine the component properties. Based on this information, we discuss the past and future evolution of V106, the constraints it puts on binary star evolution models, and the potential problems it causes for age distribution studies of MW stars.

\section{Observations and measurements}
\label{sec:obs}

\subsection{V106}

V106 was fist identified as a variable star by \citet{Mochejska2005}. They found V106 to be an eclipsing W UMa type contact system with a period of 1.4464 d. 

%Later, \citet{Tofflemire2014} determined a 99\% proper motion membership to \ngc and identified V106 as a rapidly rotating single-lined spectroscopic binary. From only three epochs of spectroscopic measurements they were unable to determine a spectroscopic membership probabillity, but mentioned that V106 (=WOCS54008) is a BSS if cluster membership is confirmed.

Later, \citet{Platais2011} measured proper motions for NGC 6791 and calculated the cluster membership probability of V106 (=WOCS~54008) to be 99\% and, thus, showed that it is a very likely cluster member. V106 is located at 3.6 arcmin from the cluster center, equivalent to $0.8R_h$ (projected half-mass radius). Visual inspection of the best photographic plate revealed a $\sim$4 magnitudes fainter star 2.7" away from V106 in the North-West direction. The proper motion of this star was not measured due to the frequent image bias by its bright and close neighbor. The measurements of radial velocities by \citet{Tofflemire2014} indicated that V106 is a rapidly rotating single-lined spectroscopic binary and assigned to it the status of a member candidate, stressing that it is a blue straggler if membership is confirmed.

\begin{figure*}
   \centering
   \includegraphics[width=18.0cm]{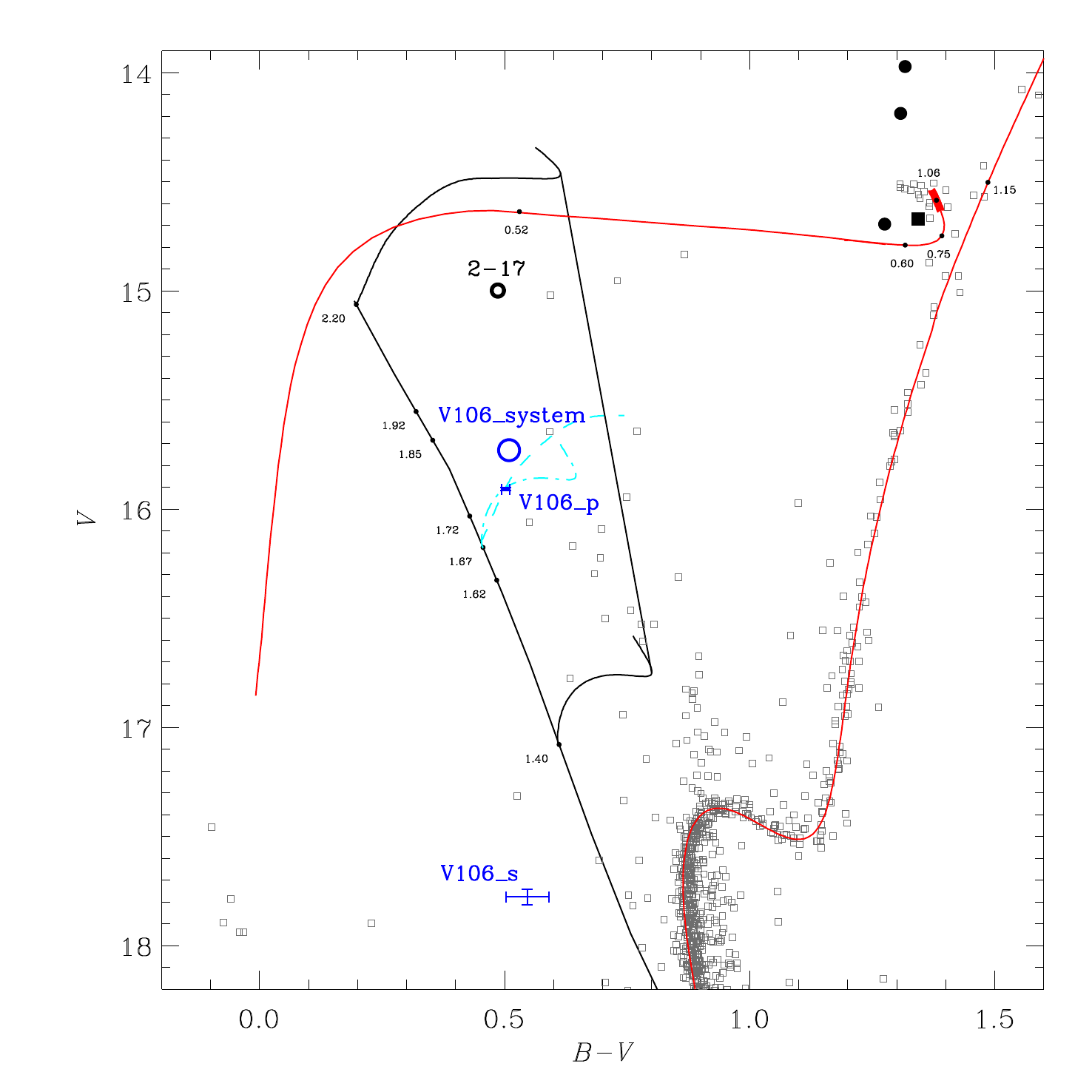}
% ### note this figure is produced from the IDL program hb_vs_bs_2017.pro ####
   \caption{Colour-magnitude diagram of \ngc with V106 and its two components along with the best matching isochrone model and ZAHB (red tracks). Small open squares are the photometry from \citet{Brogaard2012} covering the central 10 arcsec of the cluster.
The total light of V106 is shown as the blue open circle, and its primary and secondary components as blue error-boxes. The star 2-17 is shown with a black open circle. Two evolutionary tracks are shown in black, one for a $1.4 \mathrm{\rm M_{\odot}}$ star and one for a $2.2 \mathrm{\rm M_{\odot}}$ star, along with the ZAMS adopted from \citet{Brogaard2012}. The same tracks, shown in cyan, have been shifted in colour and magnitude in order to pass through the photometry of the V106 primary component. Numbers along the sequences mark the stellar mass at different locations.
Filled black circles and square mark over-massive giant cluster members \citep{Brogaard2016,Corsaro2012}.}
             \label{fig:CMD}%
    \end{figure*}

In the following, we confirm cluster membership and find that V106 is a non-eclipsing, double-lined spectroscopic binary that displays ellipsoidal variations. Identification names, coordinates and parameters are given in Table~\ref{tab:Parameters} and the location of V106 and its components in a colour-magnitude diagram (CMD) of \ngc is shown in Fig.~\ref{fig:CMD}.

%______________________________________________ 
%   \begin{figure*}
%   \centering
%   \includegraphics[width=17.0cm]{CMDwithModels2New.pdf}
%   \caption{Colour magnitude diagram of \ngc with V106 and its two components.
%V106 is shown as an orange X, its primary component as a magenta star and its secondary
%component as a cyan star. The star 2–17 is shown with a green X. Two evolutionary tracks
%are shown, one for a $1.4 \rm M_{\odot}$ star and one for a $2.2 \rm M_{\odot}$ star, and the ZAMS as used by \citet{Brogaard2012}. Interpolation between the two evolutionary tracks has been
%made to plot what corresponds to a $1.68\pm0.02 \rm M_{\odot}$ evolutionary track
%that fits well with the primary component of V106.}
%             \label{fig:CMD}%
%    \end{figure*}
%
%______________________________________________ 
   \begin{figure*}
   \centering
   \includegraphics[width=18.0cm]{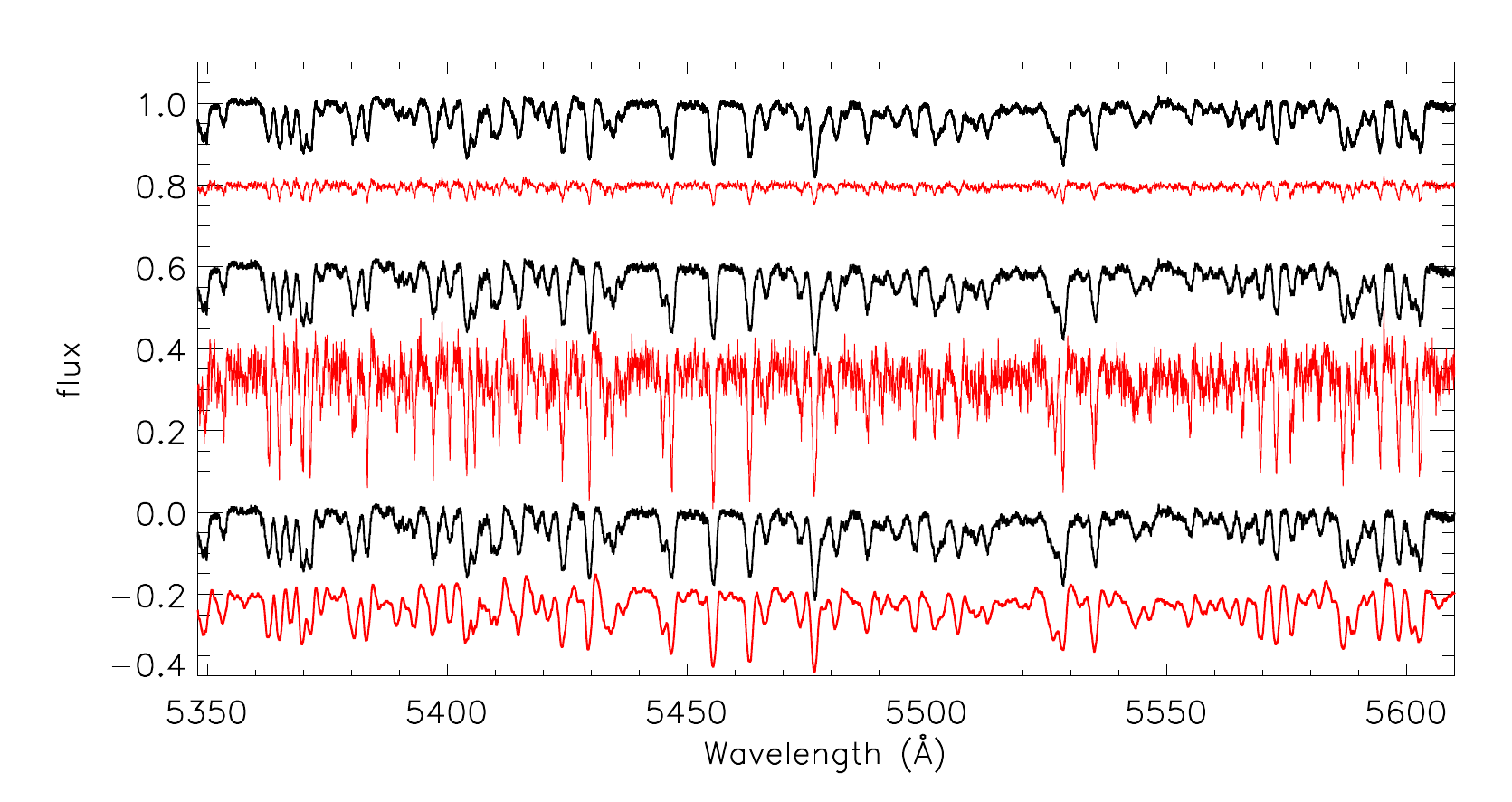}
% ### note this figure is produced from the IDL program hb_vs_bs_2017.pro ####
   \caption{The disentangled spectra of the V106 components, primary in black, secondary in red. The flux level of each spectrum has been shifted from 1 for clarity. From top to bottom, the figure shows first the disentangled spectra of the primary component and the secondary component. The middle two spectra are the disentangled spectra corrected for the light ratio, and the lower two spectra are the same again, but where the spectrum of the secondary component has been broadened to match the broadening of the primary spectrum. Notice the close similarity of the spectra.}
             \label{fig:spectra}
    \end{figure*}

\subsection{Photometry}

We used time series observations of V106 from the {\it Kepler} mission processed as described in \citet{Handberg2014} for light curve analysis. 
To estimate the orbital period and an associated uncertainty, the light curve was phased using a range of trial periods and inspected visually. The optimal period found this way corresponds also to the longest period and twice the period of the highest peak in the amplitude spectrum calculated using the program PERIOD04 \citep{Lenz2004}.
For periods differing more and more from the optimal one, the difference between the optimal period and the shifted period was adopted as the uncertainty when the shifted period no longer gave an acceptable folding of the light curve. The orbital period was found in this way to be $1.4463\pm0.0002\,\mathrm d$, and the phased light curve is shown in the upper panel of Fig.~\ref{fig:model}.

\subsection{Spectroscopy}

We obtained spectra for a number of interesting stars in \ngc from FLAMES at the Very Large Telescope using the GIRAFFE spectrograph in Medusa mode. With the HR10 setting (533.9\,nm--561.9\,nm, central wavelength $548.8 \mathrm{nm}$) the spectral resolution is $R=19800$, corresponding to 15 $\rm{km\,s^{-1}}$. We obtained 13 epochs of spectra, 11 with exposure times of 5400 seconds, the last two with exposure time of 4800 seconds. The spectra were reduced using the standard instrument pipeline. Thorium-Argon (ThAr) wavelength calibration frames were not obtained during the night of observation and the simultanous ThAr option was disabled to avoid contamination into the spectra of the faint targets. Instead, we applied radial velocity zero-point corrections to the spectra calibrated with standard day-time calibration frames. These corrections were calculated from changes in the nightly mean radial velocity of bright giant stars that were also observed. This zero-point pattern was then shifted by $-0.18\, \rm{km\,s^{-1}}$ to put the observed OI sky emission line at its expected absolute wavelength of 5577.34 \AA\, at all epochs. After this procedure, the radial velocity epoch-to-epoch RMS and half-range of single members drop significantly to about 0.15 and 0.2 $\rm{km\,s^{-1}}$, respectively. Based on these numbers, we estimate the absolute uncertainty of our radial velocity zero-point is about $0.2\,\rm{km\,s^{-1}}$.

For V106 we measured spectroscopic radial velocity (RV), projected rotational velocity $v\sin(i)$, and luminosity ratio using the broadening function (BF) formalism \citep{Rucinski2002}. Spectral lines from both components are visible in 11 epochs of the spectra making V106 a SB2 system. The measured radial velocities are given in Table~\ref{tab:rv} in the appendix. The luminosity ratio, $L_{\rm s}/L_{\rm p} = 0.179 \pm 0.007$, was determined as the ratio of the areas under the BF peak of each component at epochs where they are separated, as explained in \citet{Kaluzny2006}. Since the central wavelength is close to the center of the $V$-band, we adopt the same luminosity ratio for the $V$-band.

We used the program SBOP \citep{Etzel2004} to find an orbital solution and determine the minimum masses $M_{p,s} \sin^3 i$, projected separation $a\sin i$, and system velocity. The system radial velocity $\gamma=-44.15\pm0.18\, \rm{km\,s^{-1}}$ indicates radial velocity membership of \ngc when compared to the mean cluster velocity of \ngc, which is $-47.40\pm0.13\, \rm{km\,s^{-1}}$ with a dispersion of $1.1\, \rm{km\,s^{-1}}$ according to \citet{Tofflemire2014}. We obtained $P_{\rm RV}=58\%$ following the formal equation of \citet{Tofflemire2014} with their RV distributions of the cluster and the field stars given as gaussians. Measurement uncertainties and potential difference in the RV zeropoint between their study and ours has a significant influence on the exact number. If we reduce $\gamma$ by one sigma, then we obtain $P_{\rm RV}=69\%$. 

%Using the primary mass estimate from the CMD the secondary mass was then determined to be $M_{\rm s} = 0.221\pm0.006 \rm M_{\odot}$ along with an orbit inclination of $i=51.9\pm0.6$ degrees. 

%NOTE : It is important to remember that the calculated inclination is only an estimate since the mass of the primary component is only an estimate. Solutions to the light curve may require a different %value of the inclination. 

The radius \emph{R} of a star is related to the projected rotational velocity $v\cdot \sin(i)$ through the relation
%%%%% Radius %%%%%%%%
\begin{equation}
R = \frac{v\cdot \sin(i) \cdot P_{\rm rot}}{2\pi \cdot \sin(i)},
\label{eqn:radius}
\end{equation}
%%%
where $P_{\rm rot}$ is the rotational period of the star. The light curve of V106 shows periodic ellipsoidal variations with the orbital period from spectroscopy (see Fig.~\ref{fig:model}), and the orbital period is short. Therefore, the system is assumed to be in locked rotation with the orbital and rotational periods being equal and both stellar spin axes aligned with the orbital angular momentum vector.

%Using those assumtions, the radius of the primary component is $R_{\rm p} = 1.86\pm0.08 \rm R_{\odot}$ and for the secondary $R_{\rm s} = 1.20\pm 0.15 \rm R_{\odot}$. 

Since the system is not eclipsing, the orbit inclination is unknown, but it is still possible to find the ratio of the radii, $\frac{R_{\rm s}}{R_{\rm p}}=0.453\pm0.028$, by using Eqn.~\ref{eqn:radius} for each component with our $v\sin(i)$ measurements in Table~\ref{tab:Parameters} and taking their ratio. The uncertainty is the standard deviation of the mean from measurements on each individual spectrum.
Combining this with the light ratio and the definition of luminosity, the ratio of effective temperatures, $\frac{T_{\rm s}}{T_{\rm p}}=0.967\pm0.030$, was determined. 

We disentangled the component spectra using a spectral separation code based on the description of \citet{Gonzalez2006}, and corrected the flux levels according to the light ratio determined above. By applying an additional broadening to the spectrum of the secondary, the two spectra become very similar, as shown in Fig.~\ref{fig:spectra}. This supports our finding that the two stars have similar effective temperatures.

The spectroscopic parameters of V106 are summarized in Table~\ref{tab:Parameters}. 

\begin{table}
\centering
%\small
\caption{V106 model independent parameters.}
    \begin{tabular}{lr}
\hline
\hline
KIC ID & 2438249\\
WOCS ID & 54008 \\
        $\alpha_{\rm J2000}$ & 19 21 10.7  \\
        $\delta_{\rm J2000}$ & +37 45 31.6  \\
        $V_{\rm TOT,V106}$ & 15.7302\\
        $B_{\rm TOT,V106}$ & 16.2391\\
\hline
        Period (d) & $1.4463(2) $   \\
        $M_{\rm p}\cdot \sin^3 i$ [$\rm M_{\odot}$] & $0.837(7)$\\
        $M_{\rm s}\cdot \sin^3 i$ [$\rm M_{\odot}$] & $0.091(1)$\\
        $a\cdot \sin i$ [$\rm R_{\odot}$] & $5.244(20)$ \\
        $v\cdot \sin i_{\rm p}$ [$\rm km\cdot s^{-1}$]  & $53.0(11)$ \\
        $v\cdot \sin i_{\rm s}$ [$\rm km\cdot s^{-1}$]  & $24.0(14)$ \\
        Mass Ratio \emph{q}=$M_{\rm s}/M_{\rm p}$ & 0.109(1)  \\
        $(L_{\rm s}/L_{\rm p})_V$ & 0.179(7)  \\
\hline
$V_{\rm p}$ & 15.909(6) \\
$V_{\rm s}$ & 17.777(35) \\
$R_{\rm p}\times \sin i$ & $1.515(29)$ \\
$R_{\rm s}/R_{\rm p}$ & $0.453(28)$ \\
$T_{\rm s}/T_{\rm p}$ & $0.967(30)$ \\
$\Delta(B-V)$ & 0.04(1) \\
$B_{\rm p}-B_{\rm s}$ & 1.908(36) \\
$(L_{\rm s}/L_{\rm p})_B$ & 0.172(5)\\
$B_{\rm p}$ & 16.411(5) \\
$B_{\rm s}$ & 18.323(27) \\
$(B-V)_{\rm p}$ & 0.502(8) \\
$(B-V)_{\rm s}$ & 0.546(44) \\
$T_{\rm p}$ [K] & 7110(100)  \\
$T_{\rm s}$ [K] & 6875(200)  \\
$R_{\rm p}/\rm R_{\odot}$ from distance &  1.890(77) \\
Inclination \emph{i} [\textdegree]                   &   53.3(29) \\
$M_{\rm p}/\rm M_{\odot}$ &      1.62(21) \\
$M_{\rm s}/\rm M_{\odot}$ &      0.176(23) \\
$R_{\rm p}/\rm R_{\odot}$ &      1.890(77) \\
$R_{\rm s}/\rm R_{\odot}$ &     0.855(35) \\
Separation \emph{a} [$\rm R_{\odot}$] & 6.54(27)  \\
\hline
    \end{tabular} 
\label{tab:Parameters}
\end{table}
%%%%%%%%%%%%

\subsection{Absolute parameters}

We used the membership of \ngc to derive the photometry of the individual components of V106 and to determine the radius of the primary component. By comparing that to $R\cdot\sin(i)$, we determined the inclination, and hence the masses, radii and orbit separation of V106, as follows, yielding the component parameters given in Table~\ref{tab:Parameters}.

%The uncertainty on the magnitude of V106 is chosen to be the same as the uncertainty on the distance modulus

The apparent $B$ and $V$-magnitudes of the total system are known from the photometry of \citet{Brogaard2012} and the apparent distance modulus $(m-M)_V=13.51\pm0.06$ of \ngc was determined by \citet{Brogaard2012}. Combining this with the luminosity ratio determined above allows the derivation of the component magnitudes, $V_{\rm p}$ and $V_{\rm s}$. The apparent distance modulus of V106 for an assumed $E(B-V)=0.16$ calculated from the Gaia DR2 parallax \citep{Luri2018} is $(m-M)_V=13.72\pm0.30$ without a systematic zeropoint correction. Several investigations mentioned below have however found that the Gaia DR2 parallaxes are too small, and we investigated the potential effect of this. We obtain $(m-M)_V=13.45\pm0.30$ with the parallax zeropoint offset suggested by \citet[0.03 mas]{Luri2018}, $(m-M)_V=13.32\pm0.30$ with the offset of \citet[0.046 mas from Cepheids]{Riess2018}, and $(m-M)_V=13.27\pm0.30$ using the offset by \citet[0.0528 mas from asteroseismology]{Zinn2018}. Our own on-going investigation of eclipsing binary stars with a giant component from \citet{Brogaard2018} suggests a mean zero-point correction of 0.04 mas resulting in   
$(m-M)_V=13.37\pm0.30$.
While this is yet another strong cluster membership indication, the numbers are too uncertain to be of direct use in our analysis. For inter-comparison that suggests membership, the uncorrected Gaia DR2 parallax of V106 is $0.2262\pm0.0316$ mas, while that of the comparison star 2-17 (=KIC\,2437762, see below) is $0.1713\pm0.0250$ mas, and for three cluster member red giant branch stars KIC\,2437353, KIC\,2570094, and KIC\,2438140 the parallaxes are $0.1881\pm0.0202$, $0.2122\pm0.0247$, and $0.1817\pm0.0240$ mas, respectively.

The effective temperature of the primary can be estimated from the CMD because V106 is nearly vertically aligned with the star 2--17 investigated in detail by \citet{Brogaard2012}, see Fig.~\ref{fig:CMD}. The star 2--17 has an effective temperature of $7150$\,K \citep{Brogaard2012}, which was adopted as a first estimate for the primary component of V106. 2--17 does lie a bit further to the blue in the CMD compared to the total light of V106, but V106 consists of two components where the primary is hotter than the secondary. Using the colour-$T_{\rm eff}$ relations of \citet{Casagrande2014}, we estimated the $(B-V)$ colour difference between the two components, assuming an effective temperature of $7150\pm100$\,K for the primary, and $T_{\rm s}/T_{\rm p} = 0.967\pm0.030$ as derived above. This yielded $(B-V)_{\rm s}-(B-V)_{\rm p}=0.04\pm0.01$ for reasonable values of log$g$ close to our final values in Table ~\ref{tab:adopted}. Combined with $V_{\rm p}$ and $V_{\rm s}$ we then obtained $B_{\rm s}-B_{\rm p}=1.908\pm0.036$, $(L_{\rm s}/L_{\rm p})_B=0.172\pm0.05$, and finally the component $B$-magnitudes and the component colours $(B-V)_{\rm p}$ and $(B-V)_{\rm s}$. The latter was translated into the final $T_{\rm eff}$ estimate for the primary using the colour-$T_{\rm eff}$ relations of \citet{Casagrande2014}. Because uncertainties in reddening and metallicity, as well as their treatments in colour-$T_{\rm eff}$ relations can cause significant errors in this procedure, we adopted [Fe/H]$=+0.35$ with a nominal reddening of $E(B-V)=0.176$ to reproduce the observed $(B-V)=0.485$ colour of 2--17 for its spectroscopic $T_{\rm eff}$ \citep{Brogaard2012}. This resulted in $T_{\rm p}=7110\pm40$ K, while adopting instead a nominal reddening of $E(B-V)=0.16$, which corresponds to $E(B-V)=0.142$ at the turn-off colour \citep{VandenBerg2014,Casagrande2014} and thus consistent with the work of \citet{Brogaard2012}, gives an effective temperature 60 K cooler. A change to the assumed [Fe/H] has no effect on $T_{\rm eff}$ in the first procedure, where it would be compensated by a change in $E(B-V)$. In the case of a fixed $E(B-V)=0.16$, $\pm0.05$ dex on [Fe/H] corresponds to $\pm25$\,K on $T_{\rm eff}$. 
We adopted $T_{\rm p}=7110\pm100$ K and, from the temperature ratio derived above, $T_{\rm s}=6875\pm200$. For comparison, the Gaia DR2 effective temperature of V106, treated as a single star, is $T_{\rm eff}=7044$ K \citep{Andrae2018}.  

With the effective temperatures determined, we assumed an apparent distance modulus $(m-M)_V=13.51\pm0.06$ and inverted equation (10) of \citet{Torres2010} for the absolute magnitude to obtain the radius of the primary component $R_{\rm p}=1.890\pm0.077 \rm R_{\odot}$. The inclination then followed by evaluation against $R_{\rm p}\cdot \sin(i)=1.515\pm0.029 \rm R_{\odot}$ from Eqn.~\ref{eqn:radius}, giving $i=53.3\pm2.9$\textdegree. Using this inclination, the absolute masses and orbital separation were determined.

Since the equations for the minimum masses contain the third power of the inclination, the 2.9 degrees uncertainty causes a significant uncertainty on the absolute masses. Therefore, to more tightly constrain the masses, we also used an alternative method for their derivation. In the CMD of \ngc in Fig.~\ref{fig:CMD} we compare the photometry of V106 to a zero age main sequence (ZAMS) for the cluster parameters as determined by \cite{Brogaard2012} along with evolutionary tracks of stars with masses 1.4 and 2.2 $\mathrm{\rm M_{\odot}}$, respectively (blue lines). We shifted these evolutionary tracks along the ZAMS so they pass through the CMD position of the primary component of V106 (cyan lines). Doing so, we can see that, although the shape of the main sequence evolutionary tracks change from 1.4 to 2.2 $\rm M_{\odot}$, this has little effect on the predicted ZAMS location of V106. The predicted mass of the V106 primary is found to be 1.67 $\rm M_{\odot}$ with an uncertainty which is clearly much smaller than the previous estimate, although now dependent on the accuracy of a stellar model, and assuming that the star evolved as a single star. In our models of binary evolution, the primary star of V106 tends to be more luminous at a given mass, and direct comparisons of single and binary models show that single-star models overestimate the primary mass by about $0.05 \rm M_{\odot}$. However, since we are currently not able to reproduce the parameters of the V106 primary with our binary evolution models, we adopt this as an uncertainty rather than a shift in mass.

Comparing to the mass at different locations on the ZAMS (see Fig.~\ref{fig:CMD}) we add in quadrature an additional $\pm0.05 \rm M_{\odot}$ uncertainty on the V106 primary mass to account for uncertainties in distance, reddening and model parameters. This yields $M_p=1.67\pm0.07 \rm M_{\odot}$. With this primary mass and uncertainty, the inclination of the system becomes $i=52.59\pm1.11$ degrees when comparing to the minimum mass and the corresponding radius is $R_{\rm p}=1.908\pm0.046 \mathrm{\rm R_{\odot}}$, both in close agreement with the first estimates, but much more precise. We take these, along with those corresponding to the secondary, as our best estimates of the V106 component properties and summarise them in Table~\ref{tab:adopted}, along with the $T_{\rm eff}$ estimates which remain the same.

\begin{figure*}
   \centering
   \includegraphics[width=17.0cm]{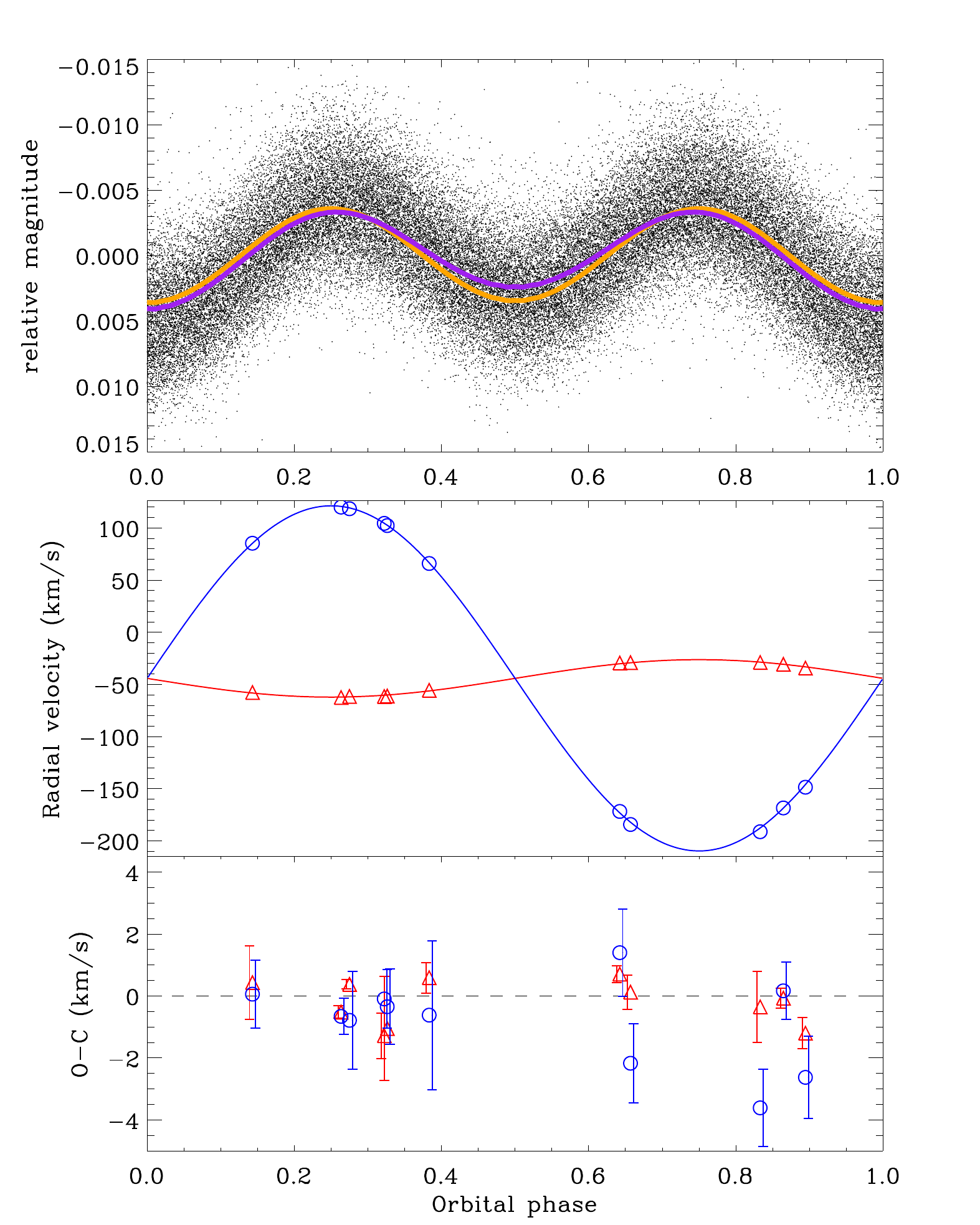}
% ### note this figure is produced from the IDL program hb_vs_bs_2017.pro ####
   \caption{Observed \textit{Kepler} light curve and radial velocity measurements of V106 compared to a PHOEBE 0.32 binary model using the parameters in Tables~\ref{tab:Parameters} and\,~\ref{tab:adopted}.  Upper panel: phased light curve with models overplotted. The orange model corresponds to the parameters inferred without the light curve, with $T_{\rm eff,s}$=6875 K. The purple model has the same parameters, except $T_{\rm eff,s}$=6500 K. Middle panel: radial velocity measurements and model. The primary component is shown with red triangles, the secondary with blue open circles. Bottom panel: radial velocity O-C diagram with the same symbols as the middle panel. Error bars have been shifted slightly to the left for the primary component and slightly to the right for the secondary component for clarity.
   }
             \label{fig:model}%
    \end{figure*}

As a final check, we compare in Fig.~\ref{fig:model} the radial velocities and the \textit{Kepler} light curve to PHOEBE 0.32 \citep{Prsa2005} eclipsing binary models.
The light curve model represented by the orange line corresponds to the binary parameters inferred in previous sections (Table~\ref{tab:Parameters} and\,~\ref{tab:adopted}), where $T_{\rm eff,s}=6875\,\mathrm{K}$. This model captures the effects of ellipsoidal deformation fairly well, and serves as a sanity check. However, since the system is not eclipsing, there are degenerate solutions, and the inclination is not well constrained from the light curve alone. Eclipses begin to appear in the models at an inclination of 65.5 degrees, at which point the primary mass is $\sim 1.1 M_{\odot}$, as low as the mass of the cluster turn-off.
The difference in maximum magnitude (minimum luminosity) between phases 0.0 and 0.5 is due to local heating effects on the V106 components where they face each other. The orange model does not show this difference, whereas the purple model, which has the same parameters, except $T_{\rm eff,s}$=6500 K does. A model with $T_{\rm eff,s} > T_{\rm eff,p}$ would show the opposite effect of the observations, having the largest magnitude at phase 0.5 instead of 0.0. While there are again degenerate solutions with correlations between the radius ratio and the temperature ratio, this does confirm that the secondary star is cooler than the primary.

%______________________________________________ 

\begin{table}
\centering
%\small
\caption{V106 model dependent parameters.}
    \begin{tabular}{lrr}
\hline
\hline
Inclination $i$ [degrees] & $52.59(1.11)$ & \\
Separation $a [\rm R_{\odot}]$ & $6.602(98)$ & \\  
\hline
Parameter & Primary & Secondary \\
\hline
$M [\rm M_{\odot}]$ & 1.67(7) & 0.182(8) \\
$R [\rm R_{\odot}]$ & 1.908(46) &  0.864(57) \\
log$g$ & 4.10(1) & 3.82(7) \\
$T_{\rm eff}$ [K] & 7110(100) & 6875(200) \\
\hline
    \end{tabular} 
\label{tab:adopted}
\end{table}
%%%%%%%%%%%%

%%%%%%%%%%%%%%%%%%%%%%%%%%%%%%%%%%%%%%%%%%%%%%%%%%%%%%%%%%%%%
\section{Properties and evolution of V106}
When considering the properties of V106 and the proper motion and radial velocity membership to \ngc, it seems clear that V106 is a binary blue straggler cluster member. The current primary star is much more massive than the cluster turn-off, $M_{\rm turn-off, now}=1.09 \rm M_{\odot}$ \citep{Grundahl2008, Brogaard2011, Brogaard2012} and is also bluer and more luminous (see Fig.~\ref{fig:CMD}). The current secondary star is both oversized and overluminous for its mass under the (false) assumption that it is a main sequence star. 
From a closer inspection, however, it is immediately clear that the properties of the secondary star resembles those of a bloated (or proto) extremely low-mass helium white dwarf (proto-ELM~WD). Such an object is produced from a low-mass star in a close binary which loses its hydrogen-rich envelope via Roche-lobe overflow (RLO) to its companion star, thereby exposing its (almost) naked degenerate helium core \citep[e.g.][and references therein]{itla14}. Furthermore, the orbital period ($P=1.45\,{\rm d}$) and the secondary mass ($M_2=0.182\,\rm M_{\odot}$) also fit perfect with the orbital period--mass correlation for binary helium WDs \citep[see e.g. Figure~4a in][]{ts99}.

It is not surprising that V106 is a BSS+WD binary. In the open cluster NGC\,188, \citet{gmg+15} found evidence for 14 BSS binaries which also formed via mass transfer. Another example is the eclipsing binary KIC~8262223 \citep{ggm+17} which contains a $\delta$~Scuti pulsator and a $0.20\,\rm M_{\odot}$ proto-ELM~WD with an orbital period of $P=1.61\,{\rm d}$, i.e. a system somewhat resembling V106.

Knowing that the current primary star must have gained a significant amount of mass, the current secondary must have lost at least the same amount of mass. Thus, the only reasonable interpretation is that the current secondary star started out as the most massive component. As it evolved through the Hertzsprung gap it overfilled its Roche lobe and mass transfer to the companion began. 
Since we know the current total mass of the system as well as the turn-off mass of the cluster, we can determine the minimum original mass of the current primary star, which started out as the least massive star. If we assume no mass-loss during mass-transfer, the calculation is $M_{\rm min, primary, then}=M_{\rm system,now}-M_{\rm secondary, then}=M_{\rm system,now}-M_{\rm turn-off,then}=1.85\,\rm M_{\odot}-1.15\,\rm M_{\odot}= 0.70\,\rm M_{\odot}$. Here, primary and secondary refer to the current components, and we have used the mass of current red giants in \ngc (see Fig.~\ref{fig:CMD} and \citealt{Brogaard2012}) for the turn-off mass to allow V106 to have existed in its current form for some time, as suggested by the position of the primary component in the CMD. The maximum amount of mass that can have been lost from the system during the mass transfer phase is $M_{\rm lost, max}=M_{\rm max, system, then}-M_{\rm system,now}=2\times M_{\rm turn-off, then}-M_{\rm system, now} = 2\times1.15\,\rm M_{\odot} - 1.85\,\rm M_{\odot} = 0.45\,\rm M_{\odot}$ even if it started out with a mass ratio of $q=1$. This, and the short orbital period, puts tight constraints on the evolution of the system.

\subsection{Detailed binary stellar evolution modelling}
We applied the MESA code \citep[Modules for Experiments in Stellar Astrophysics, version~9793;][]{pbd+11,pca+13,pms+15,psb+18} in binary mode evolving both stars\footnote{MESA inlists are available upon request to the main author.} for calculating the evolution leading to the formation of V106.
The stars were chosen with a chemical composition of $X=0.6933$ and $Z=0.0204$ and we included orbital angular momentum changes due to magnetic braking, mass transfer/loss and gravitational wave radiation. For the mass loss from the system, we applied the isotropic re-emission model \citep{tv06}. 
The initial binary needs to be chosen carefully such that our computations can reproduce the observables of V106 (age and orbital period, as well as mass, radius and surface temperature of both stars).

The ZAMS mass of the progenitor of the current $0.182\,\rm M_{\odot}$ secondary star (the proto-ELM~WD) was chosen to be $1.15\,\rm M_{\odot}$ in order to produce a final age of V106 in accordance with the cluster age. The ZAMS mass of its companion star (the progenitor of the current $1.67\,\rm M_{\odot}$ BSS) is taken to be $0.80\,\rm M_{\odot}$. A somewhat challenging constraint to model for this close-orbit binary system is that the current BSS does not fill its Roche lobe. Hence, any simulated binary must avoid that the accreting star (producing the BSS) evolves to a (sub)giant stage, and initiates RLO back to the donor star, before the original donor star terminates its RLO. In other words, the progenitor star of the proto-ELM~WD must be able to completely terminate its RLO before the BSS evolves to become a (sub)giant. This constraint puts a tight limit on the combination of the ZAMS masses of the two progenitor stars and the mass loss from the system during RLO. 

Our best solution for a progenitor system which evolves to reproduce V106 consists of an initial ZAMS binary with two stars of masses $1.15\,\rm M_{\odot}$ and $0.80\,\rm M_{\odot}$, and an orbital period of $3.42\,{\rm d}$. The magnetic braking mechanism \citep{vz81,rvj83} is treated following \citet{itl14} using an ad~hoc value of $\gamma = 4$. The value of $\gamma$ is not well constrained, nor is the exact prescription of the orbital angular momentum loss \citep[e.g.][]{vvp05}, let alone the magnetic wind \citep{gdd18}. The exact value of $\gamma$, however, is not important here since applying a different strength of the orbital angular momentum loss will simply translate into a larger or smaller value of the initial orbital period to reproduce the same system \citep{itl14}.
For a fixed value of $\gamma$, trial-and-error fine-tuning of the initial orbital period is needed to produce the desired final orbital period equal to that of V106 ($P=1.45~\rm{d}$). If the strength of magnetic braking is weaker than assumed here, the initial orbital period of the progenitor system would need to be smaller to reproduce V106. For example, for a fixed initial orbital period of 3.42~d on the ZAMS, applying $\gamma=2$, 4 or 5, results in onset of RLO at orbital periods of 2.45~d, 0.899~d and 0.727~d, at stellar ages of 6.97~Gyr, 6.35~Gyr and 5.77~Gyr, respectively.
 
\begin{figure}
\centering
\includegraphics[angle=-90,width=8.5cm]{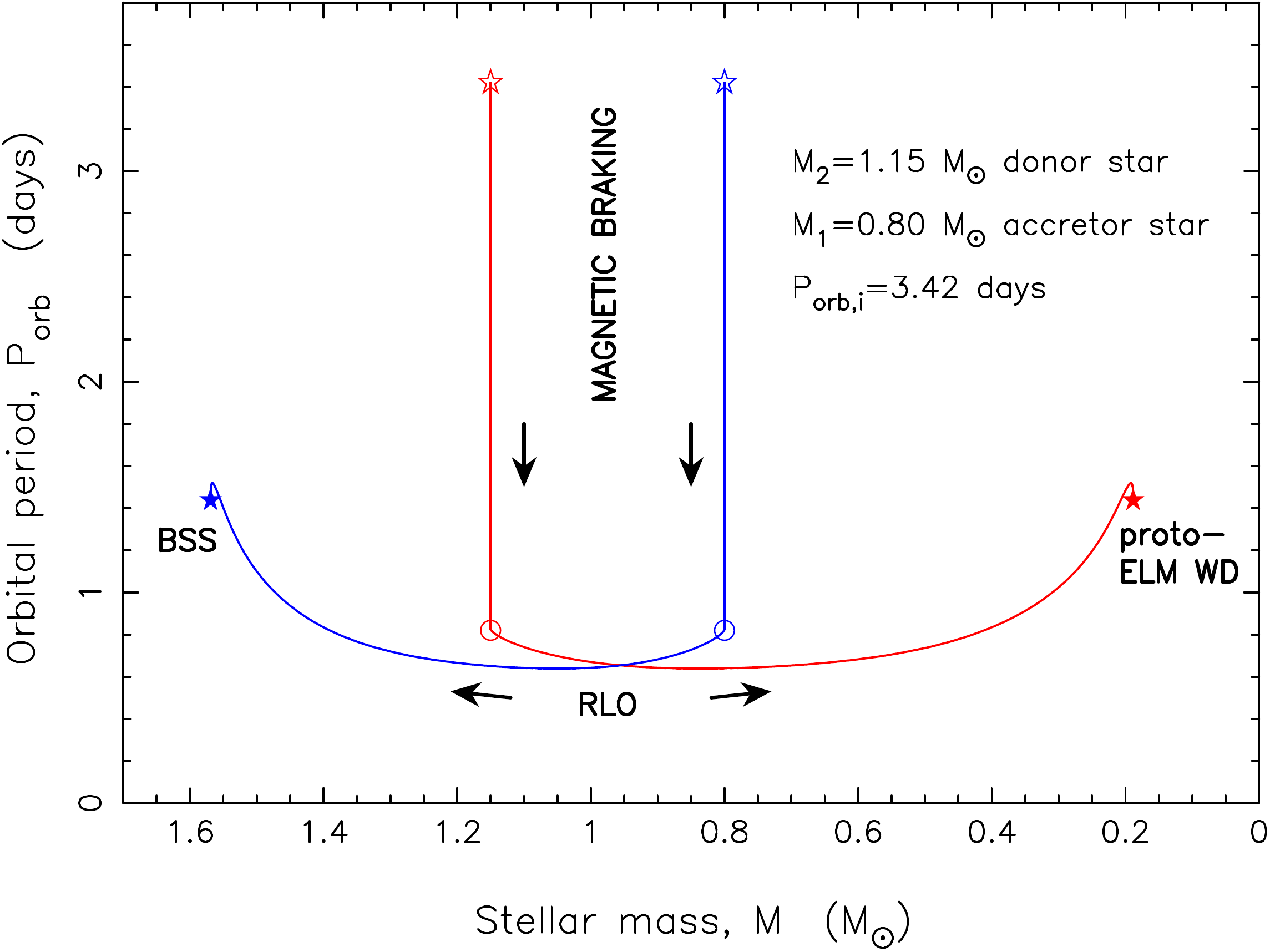}
\caption{
Formation model of V106 via mass transfer in a binary system. The initial configuration is a binary with two ZAMS stars (open star symbols) of masses $1.15$ and $0.80\;\rm M_{\odot}$, in a circular orbit with an orbital period of $3.42\;{\rm d}$. The $1.15\;\rm M_{\odot}$ donor star evolves to become the present secondary star ($\sim\!0.18\;\rm M_{\odot}$ proto-ELM~WD), while the $0.80\;\rm M_{\odot}$ accretes material and produces the present primary star ($\sim\!1.6\;\rm M_{\odot}$ BSS) -- see solid star symbols. As a result of magnetic braking, the orbital period decreases to about $0.82\;{\rm d}$ prior to RLO (open circles). See text for details.
}
\label{fig:m2m1Porb}
\end{figure}

The orbital period evolution of the progenitor system of V106, as a function of the masses of the stellar components, is shown in Fig.~\ref{fig:m2m1Porb}. 
The magnetic braking leads to loss of orbital angular momentum such that the orbital period decreases to $P\simeq 0.82\,{\rm d}$ at the onset of RLO. The mass transfer is initiated shortly after core hydrogen exhaustion, i.e. early Case~B RLO \citep{kw67}. During RLO, the orbital period initially shrinks further until it starts to increase shortly after the mass ratio reversal. Disregarding the formation of a circumbinary disk and assuming that the wind-mass loss rate from the donor star is negligible compared to the rate at which material is transfered via the first Lagrangian point during RLO, the isotropic re-emission model can simply be described by the parameter $\beta$. This parameter (assumed to be constant) is the fraction of transfered material which is ejected from the accretor, and thus carrying the specific orbital angular momentum of the accretor. Such mass loss from the system ($\beta \neq 0$) is expected if either the mass-transfer rate is high and/or the accreting star evolves close to critical rotation. The latter effect is particular important in high-mass binaries where $\beta$ values up to 0.90 are possible \citep{plv05}. In our modelling of a low-mass binary, we assume that 20~per~cent of the transferred material is lost from the system, i.e. $\beta =0.20$ (although this ad~hoc value is quite uncertain and could easily be, e.g., 5 or 25~per~cent). 
Assuming a constant value for $\beta$ during RLO, the change in orbital separation is given by \citep{tau96}:
\begin{equation}\label{eq:tau96}
  \frac{a}{a_0}= \left( \frac{q_0(1-\beta)+1}{q(1-\beta)+1} \right) ^{\frac{3\beta-5}{1-\beta}}
                 \left( \frac{q_0+1}{q+1} \right) \left( \frac{q_0}{q} \right) ^2,
\end{equation}
where $a_0$ and $a$ refer to the orbital separations before and during RLO, respectively, and where $q_0$ and $q$ represent the
mass ratios at these two epochs.

\begin{figure}
\centering
\includegraphics[angle=-90,width=8.5cm]{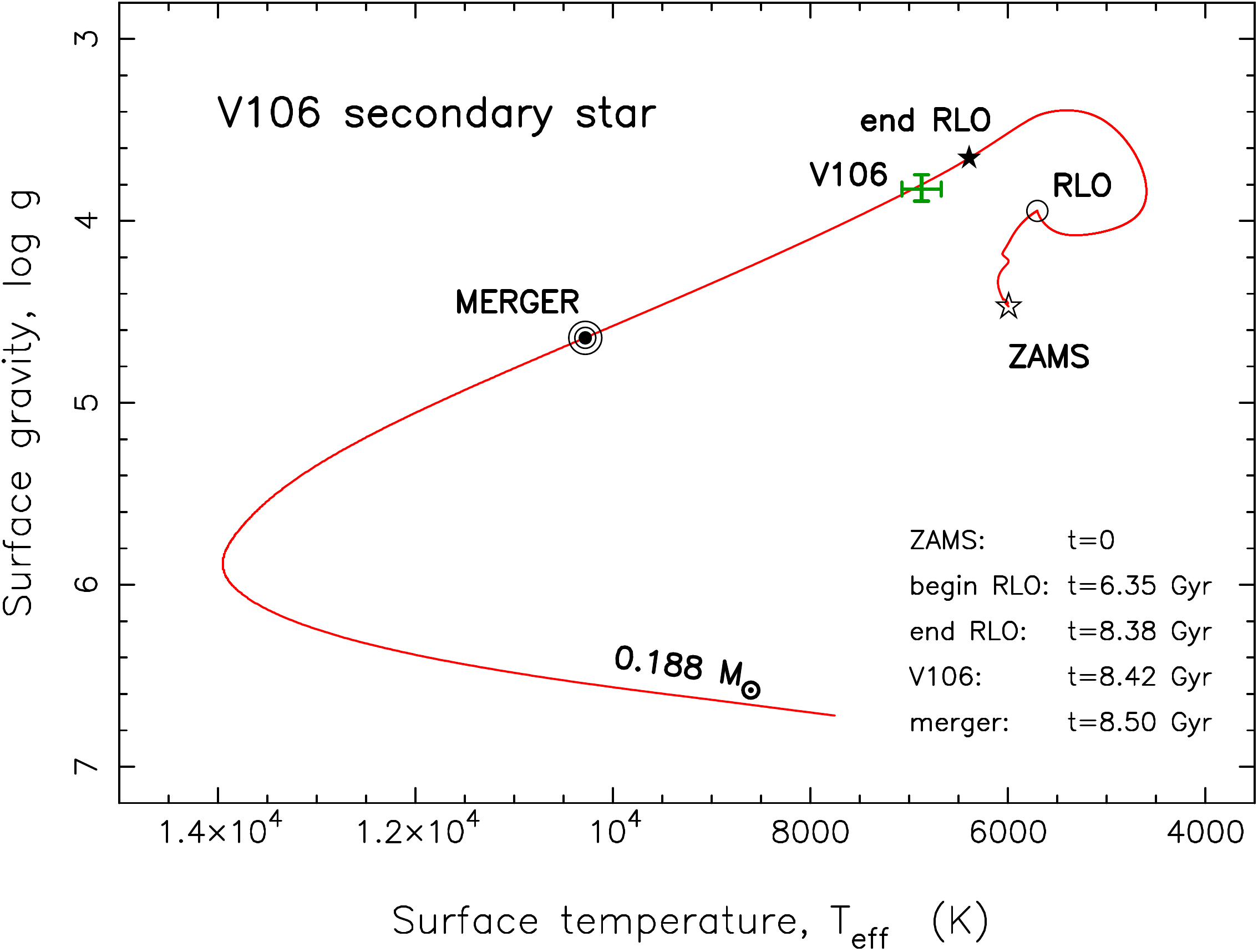}
\caption{
Past and future evolution of the secondary star in V106 in the ($T_{\rm eff},\log\,g$)~diagram, according to our model.
The current location of the observed $0.182\pm0.006\;\rm M_{\odot}$ secondary star in V106 is plotted with a green cross (error bars shown) on top of our track for a $0.188\,\rm M_{\odot}$ proto-ELM~WD.
V106 is expected to initiate a merger event in about $80\;{\rm Myr}$.
}
\label{fig:m2_Teff_logg}
\end{figure}
\begin{figure}
\centering
\includegraphics[angle=-90,width=8.5cm]{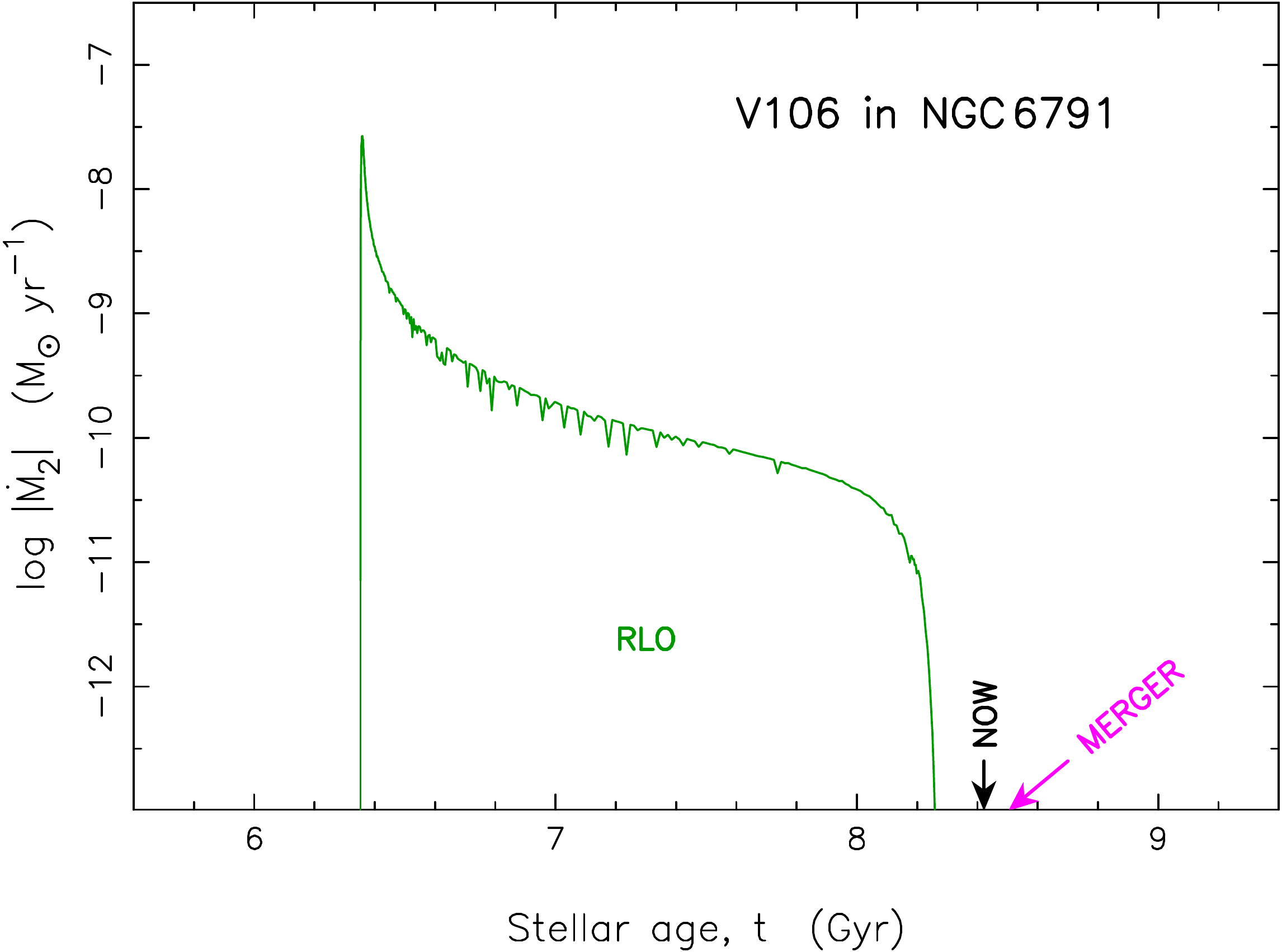}
\caption{
Mass-transfer rate as a function of stellar age of the former donor star in V106 (see Fig.~\ref{fig:m2m1Porb}). The RLO ceased completely at a stellar age of about $8.38\;{\rm Gyr}$, some $\sim\!40\;{\rm Myr}$ ago. The present age of the modelled system is $t=8.42\;{\rm Gyr}$ (marked by \textsc{"now"}), in agreement with current age estimates of \ngc. It is anticipated that the system will merge in about $80\;{\rm Myr}$ when the primary star (BSS) fills its Roche lobe (see text). Some numerical noise is seen in the calculated values of $|\dot{M}_2|$.
}
\label{fig:m2dot}
\end{figure}
\begin{figure}
\centering
\includegraphics[angle=-90,width=8.5cm]{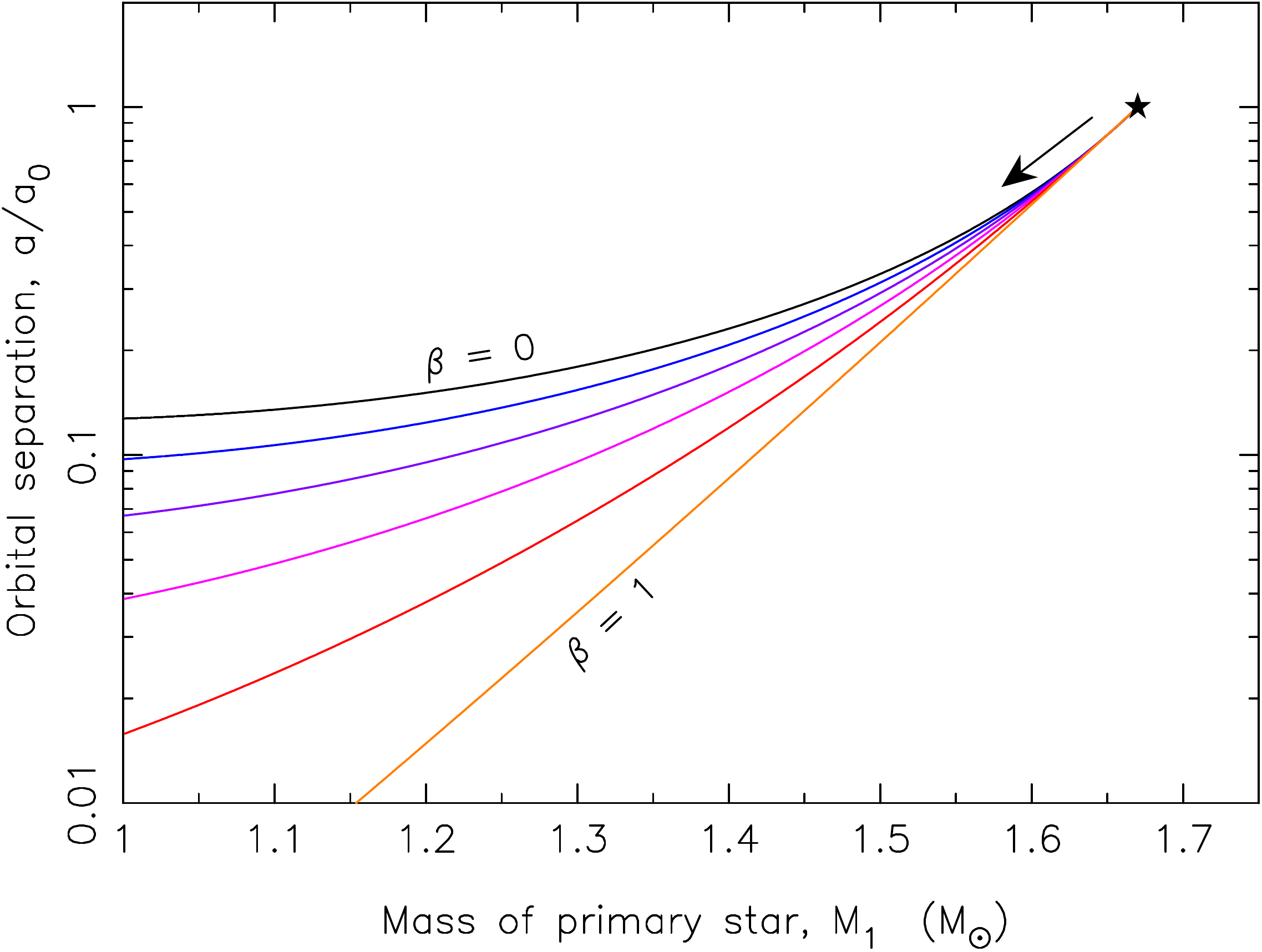}
\caption{
Future orbital evolution of V106 once the primary star initiates RLO. The calculations are based on the isotropic re-emission model (see Eq.~\ref{eq:tau96}) with different values of $\beta$ in steps of 0.2. The plot shows the decrease in orbital separation, $a$ (in units of the pre-RLO orbital separation, $a_0$) as a function of the decreasing primary star mass (evolution from right to left). The calculations are based on our derived values of a primary star mass of $1.67\;\rm M_{\odot}$ and a secondary star mass of $0.182\;\rm M_{\odot}$.
The efficient decay of the orbital separation after RLO will lead to a common-envelope event in which the secondary star will undergo in-spiral and merge with the core of the primary star.
}
\label{fig:aratio}
\end{figure}

The final post-RLO masses of the two stars from our stellar evolution model are $0.188\,\rm M_{\odot}$ and $1.57\,\rm M_{\odot}$ for the proto-ELM~WD and the BSS, respectively, whereas the observational data analysis for V106 yields $0.182\pm0.006\,\rm M_{\odot}$ and $1.67\pm0.07\,\rm M_{\odot}$. The final orbital period from our model is $1.437\,{\rm d}$\footnote{The final decrease in orbital period (i.e. the little hook shown on the tracks in Fig.~\ref{fig:m2m1Porb} near the termination of RLO) is caused by magnetic braking still assumed to be active. This may not be the case in reality.} and very close to the observational value of $1.446\,{\rm d}$. These numbers are thus in excellent agreement, except the mass of the primary star which is slightly smaller (but still within $2\sigma$) in our calculated model. The reason is that in our calculations the BSS (the accretor) will evolve too fast and become a (sub)giant before the termination of the RLO, leading to reverse mass transfer, if it has a mass above $\sim\!1.60\,\rm M_{\odot}$. 
It should be noted that our primary star model mass estimate for V106 ($1.57\;M_{\odot}$) also has some uncertainty, and that our observed value is based on a single star evolution model, and therefore our model mass is likely to be compatible within $1\sigma$.

In Fig.~\ref{fig:m2_Teff_logg}, we show the formation and further evolution of the secondary star in V106 according to our MESA model. The observed data is shown with a green cross and our computed model is the red track. After detachment from the RLO, the donor star (the proto-ELM WD) is too bloated ($R=1.15\,\rm R_{\odot}$) and too cool ($T_{\rm eff}=6204\;{\rm K}$) to fit the derived radius and temperature of the secondary star in V106. However, after some $40\,{\rm Myr}$ the radius has decreased to $R=0.870\,\rm R_{\odot}$ while the surface temperature has increased to $T_{\rm eff}=6995\,{\rm K}$. These values are also in fine agreement with our empirical values of $R=0.864\pm0.057\,\rm R_{\odot}$ and $T_{\rm eff}=6875\pm200\,{\rm K}$ derived from the observational data (Section~\ref{sec:obs}). 
As a result of the BSS becoming significantly more massive than the ZAMS mass of the initially most massive star ($M=1.15\,\rm M_{\odot}$), the BSS evolves fast and already becomes a subgiant, filling its Roche lobe (at a radius of $R=3.64\,\rm R_{\odot}$) about $120\,{\rm Myr}$ after RLO detachment ($80\,{\rm Myr}$ from now). This point is marked as the onset of the merger event (Section~\ref{subsec:future}).

The mass-transfer rate of the donor star as a function of stellar age is plotted in Fig.~\ref{fig:m2dot}. At the very onset of the RLO, the mass-transfer rate reaches almost $|\dot{M}_2|=10^{-7.5}\,\rm M_{\odot}\,{\rm yr}^{-1}$. Thus, at the early stage of the RLO, the mass-transfer timescale ($M_2/|\dot{M}_2|=25\,{\rm Myr}$) is smaller than the thermal timescale of the accreting star ($\tau_{\rm th}=89\,{\rm Myr}$). This may cause it to expand and will further justify our assumption of evolution with some mass loss from the system.

The total age of V106 according to our model is $t=8.42\,{\rm Gyr}$, which is close to the value estimated for \ngc , $8.3\pm0.3$ Gyr \citep{Brogaard2012}. If we increase the initial mass of the ZAMS progenitor of the secondary star from $1.15\,\rm M_{\odot}$ to $1.20\,\rm M_{\odot}$, or $1.25\,\rm M_{\odot}$, the age of our model would only be $t=7.6\,{\rm Gyr}$, or $t=6.7\,{\rm Gyr}$ at the time the system has detached from RLO and the secondary star has decreased its radius to the observed value. 

Whereas we can reproduce age, orbital period, $M$, $R$ and $T_{\rm eff}$ of the secondary star in V106 (Table~\ref{tab:model}), we are not successful in modelling $R$ and $T_{\rm eff}$ of the primary (accreting) star. 
The observed values are $R=1.91\,\rm R_{\odot}$ and $T_{\rm eff}=7110\,{\rm K}$, whereas our model star is fainter, larger and cooler with $R=2.87\,\rm R_{\odot}$ and $T_{\rm eff}=5489\,{\rm K}$. Although our measured $T_{\rm eff}$ is affected by uncertainties in reddening and metallicity, very similar values were found from ground based broad band photometry and Gaia DR2 with independent reddening estimates. The agreement with the star 2-17, and the very similar spectra of the primary and secondary component seen in Fig.~\ref{fig:spectra} also support our measured $T_{\rm eff}$. Furthermore, the primary $T_{\rm eff}$ of our binary evolution model is cooler than the cluster turn-off \citep{Brogaard2011} which is clearly not real according to the observed CMD. We are therefore confident that the discrepancies in the parameters of the primary star are because the evolution of the primary in our models is too fast.
In an attempt to account for this discrepancy, we performed a number of trial calculations with a larger initial helium content of $Y=0.30$ (keeping $Z$ unchanged). Indeed, this leads to a more compact and hotter primary star after accretion. We can then reproduce $R_1$ and $T_{\rm eff,1}$, but not at the correct time. The nuclear evolution timescale of the primary star is shorter and thus it becomes a subgiant and initiates RLO back to the secondary star, before the latter star terminates its mass transfer completely. This problem can be fixed by slightly increasing the initial orbital period, but then the final orbital period (and thus $M_2$) becomes too large. It is also possible to adjust some of the assumed input physics behind the rejuvenation of the primary star as it accretes material (e.g. the mixing of hydrogen into the core region of the accretor which affects the nuclear burning rate) or perhaps even the mass-transfer rate itself. In any case, what seems to be needed is a process that slows down or delays the mass increase of the primary star. 
A full investigation of this issue is beyond the scope of this paper. Nevertheless, we have demonstrated a reasonable solution and we are fairly optimistic in being able to reproduce a system like V106 from further finetuning of input parameters.

\begin{table}
\centering
\caption{V106 parameters. Comparison of values derived from a fit to the observations (second column, cf. Section~2) vs. our MESA model of binary evolution (third column).}
\begin{tabular}{lrr}
\hline
\hline
Parameter & Observations & Model \\
\hline
Age, $t$ [${\rm Gyr}$]                 &  8.3(3) & 8.42 \\
Orbital period, $P$ [${\rm d}$]     &  1.4463(2) & 1.437 \\
Secondary mass, $M_2$ [$\rm M_{\odot}$]    & 0.182(6)   & 0.188 \\
Secondary radius, $R_2$ [$\rm R_{\odot}$]  & 0.864(57)  & 0.870 \\
Secondary temperature, $T_{\rm eff,2}$ [${\rm K}$]  & 6875(200) & 6995 \\
Primary mass, $M_1$ [$\rm M_{\odot}$] & 1.67(5) & 1.57 \\
Primary radius, $R_1$ [$\rm R_{\odot}$]    & 1.908(41)  & 2.87 \\
Primary temperature, $T_{\rm eff,1}$ [${\rm K}$] & 7110(100) & 5489 \\
\hline
    \end{tabular} 
\label{tab:model}
\end{table}

\subsection{The future destiny of V106}\label{subsec:future}

As we discussed above, assuming our theoretical model to be approximately correct, the current primary star of V106 will initiate mass transfer towards the proto-ELM WD in about $80\,{\rm Myr}$. However, upon RLO the orbital separation will decrease significantly (see Fig.~\ref{fig:aratio}) because of the small mass ratio between the donor star and the accretor star, $q=0.182/1.67=0.109$. This situation will lead to a runaway mass-transfer phase, shrinking the orbit further until the WD is captured inside the envelope of the primary star, and a common envelope is formed \citep{ijc+13}. However, because at this stage the total binding energy of the envelope of the primary star (now with a radius of $R=3.64\,\rm R_{\odot}$) is still quite large \citep[e.g.][]{dt00,xl10,lvk11}, the subsequent in-spiral of the WD, as a result of frictional forces, will not be successful in ejecting the envelope. 

Integrating through the envelope of our model primary star (assuming a core boundary located at $X=0.10$), we find a gravitational binding energy of the envelope of $E_{\rm bind}=-3.14\times 10^{48}\;{\rm erg}$. However, even if the WD would continue its in-spiral to the critical innermost stable point where the primary core would fill its Roche lobe, the orbital energy at that point ($E_{\rm orb}=-7.32\times 10^{46}\;{\rm erg}$) is smaller by more than a factor of 40. Hence, even if this orbital energy could be released and converted efficiently to kinetic energy in the hydrogen envelope, the common envelope cannot be ejected from this in-spiral. Instead, a complete merger process between the two stellar objects will occur as the WD sinks to the core of the primary star.

The future of this merged BSS is not completely clear. It could be either a single giant star, which has a mass compatible with the overmassive giants found in the cluster \citep{Brogaard2012,Corsaro2012,Brogaard2016}, or the system might evolve (over time) into an extreme horizontal branch star like the ones also already known in \ngc.
However, as argued above, our MESA model calculations strongly suggest that the outcome is a single over-massive giant. We obtained the same result by independently modelling the evolution of V106 from the current parameters using the binary\_c online tool\footnote{http://personal.ph.surrey.ac.uk/cgi-bin/binary5.cgi}, which is a front-end for the population nucleosynthesis code of \citet{Izzard2004, Izzard2006, Izzard2009} based on the binary star evolution code of \citet{Hurley2002}. Recent work on the topic \citep{zhjb17} also supports our conclusion that the outcome of the future V106 merger is a normal horizontal branch star, and not a hot subdwarf. The system is possibly producing a so-called early-type R~star \citep[see Fig.2a in][]{ijl07}.
In any case, we expect that the final destiny is a CO~WD at the end of its nuclear evolution.

%Due to the short orbital period (P=1.4463 days) another mass-transfer phase is unavoidable once the current primary star evolves towards the red giant phase. Depending on the mass-transfer details V106 will evolve through a common-envelope phase to become either an overmassive giant or one of the extreme horizontal branch stars that \ngc is known to host.

%{\color{red}
%We modeled the evolution of V106 from the current parameters using the binary\_c online tool\footnote{http://www.ast.cam.ac.uk/$\sim$rgi/cgi-bin/binary5.cgi}, which is a front-end for the population nucleosynthesis code of \citet{Izzard2004, Izzard2006, Izzard2009} based on the binary star evolution code of \citet{Hurley2002}. Using the standard parameters, the system merges into a single giant star, which has a mass compatible with the overmassive giants found in the cluster \citep{Brogaard2012,Corsaro2012,Brogaard2016}.
%}

%\section{EHB and RC}

%EHB $\Delta P$ 241.3, 234.6 and 239.7 \citep{Reed2012}.
%RC 245.2 -- 313.78 \citep{Bossini2017}.

\section{Summary, conclusions and outlook}
\label{sec:conclusion}
We determined the properties of V106 and its components. V106 is a non-eclipsing, SB2 binary system that shows the effects of ellipsoidal variations and reflection. Proper motion, system radial velocity, and distance all point to membership of the open cluster \ngc. The location in the CMD therefore suggests that V106 is a blue straggler with both components bluer than the turn-off. This is supported by the atypical properties of the secondary, which must be the result of a past mass-transfer event in an Algol-type binary, leaving this star as the core remnant of what started out as the most massive component of the system. We identify the secondary star as a proto-ELM~WD. Such objects can remain bloated for up to a few Gyr as a result of continued hydrogen burning in their residual $\sim\!0.01\,M_{\odot}$ envelope \citep[][]{itla14} until they finally settle on the WD cooling track. 

BSS+WD binaries like V106 have been shown to be common in some open clusters like NGC~188 \citep{gmg+14,gmg+15}. In addition, there is a similarity between V106 and KIC~8262223, an eclipsing $\delta$~Scuti pulsator in a binary with a proto-ELM~WD \citep{ggm+17}, and also the R~CMa-type eclipsing binary KIC~6206751 \citep{lp18}. Finally, we note the semi-detached Algol system V228 in 47~Tuc \citep{Kaluzny2007}, which is a system with a $1.51\,\rm M_{\odot}$ primary star with a $0.20\,\rm M_{\odot}$ Roche-lobe filling secondary star and an orbital period of $1.15\,{\rm d}$. According to our numerical modelling, V228 is a metal-poor version of a precursor to V106 just prior to detachment of the secondary star (i.e. before the initially most massive star in the original ZAMS binary terminates RLO).

To verify our formation scenario, we have modeled the formation of V106 using the binary star extension in MESA. We find that V106 started out about $8.4\,{\rm Gyr}$ ago as a ZAMS binary with component masses of about $1.15\,\rm M_{\odot}$ and $0.80\,\rm M_{\odot}$ and an orbital period close to $3.4\,{\rm d}$. The latter value is dependent on the strength of the magnetic braking. The system detached from RLO about $40\,{\rm Myr}$ ago, producing a $\sim 0.182\;M_{\odot}$ low-mass proto-WD orbiting a $\sim 1.67\;M_{\odot}$ BSS. Whereas our numerical model can reproduce most of the observed parameters of V106 (Table~\ref{tab:model}), and thus solve for the classic Algol paradox, we fail to reproduce the radius and effective temperature of the BSS. The reason for this is possibly related to the applied input physics of the rejuvenation of the BSS during mass accretion.

Although we believe our model is correct in large terms, many details are still not accounted for. One caveat is that our modelling does not include rotation of the stars. Depending on the additional mixing and tidal forces at work, this may affect differential rotation in the envelope of both stars and thus, for example, the rejuvenation of the accreting star (BSS) as well as having a small effect on the precise radial contraction timescale of the proto-ELM~WD \citep{imt+16}. Furthermore, the proto-ELM~WD (the secondary star in V106) is not expected to rotate in synchronization with the orbit. After Roche-lobe detachment, it may spin up significantly from the fall-back of the remaining $\sim 0.01\;M_{\odot}$ H-rich envelope. This will affect our derivation of the radii of the stellar components in Section~\ref{sec:obs}. If the outer layers of the secondary star are indeed rotating faster than synchronous, then by Eq. 1, the radius is smaller than derived. However, if the radius is smaller, then the heating effects shown in the light curve can only be reproduced by a smaller $T_{\rm eff,s}$, while, at the same time, the spectroscopic light ratio requires a larger $T_{\rm eff,s}$. This suggests that the radius of the secondary cannot be much different than inferred.

As a possible explanation, loss of spin angular momentum due to strong winds of a proto-WD \citep{spr98} may counteract and limit the expected spin-up effect from Roche-lobe detachment.

The short orbital period, and hence relatively small orbital separation, means that another mass-transfer event will be unavoidable in the future when the current primary component evolves to become a giant.
According to our numerical modelling of V106, this binary will merge in about $80\,{\rm Myr}$ when the primary star fills its Roche lobe, forming a single giant star, which has a mass compatible with the over-massive giants found in the cluster \citep{Brogaard2012,Corsaro2012,Brogaard2016},
in agreement with recent work \citep{zhjb17}, which suggests that the outcome will be a normal horizontal branch star. The high mass of this giant will make it appear young for its true age, which is only revealed by the parent cluster. If found in the field, such a star could be mistaken for a young star. Therefore, V106 is likely a prototype progenitor of old field giants masquerading as young, such as discussed by \citet{Izzard2018}.

%or the system to might evolve into an extreme horizontal branch star like the ones known in \ngc.

%In any case, V106 constitutes a very interesting opportunity for obtaining new detailed information on binary evolution and the exotic stars in \ngc. 

%{\color{red}We modeled the evolution of V106 from the current parameters using the binary\_c online tool. Using standard parameters, the system evolves into a single giant star, which has a mass compatible with the overmassive giants found in the cluster \citep{Brogaard2012,Corsaro2012,Brogaard2016}.}

\section*{Acknowledgements}
We thank the anonymous referee for useful comments and suggestions that helped improve the manuscript. We thank Simon Jeffery for useful conversations. Funding for the Stellar Astrophysics Centre is provided by The Danish National Research Foundation (Grant DNRF106). RGI thanks the STFC for funding Rutherford grant ST/M003892/1. AM acknowledges the International Space Science Institute (ISSI) for the support provided to the asterosSTEP ISSI International Team. AM acknowledges support from the ERC Consolidator Grant funding scheme (project ASTEROCHRONOMETRY, G.A. n. 772293).

\bibliographystyle{mnras}
\bibliography{brogaardV106}

%%%%%%%%%%%%%%%%% APPENDICES %%%%%%%%%%%%%%%%%%%%%

\appendix

\section{RV measurements}

\begin{table}
\centering
%\small
\caption{Individual RV measurements of V106}
    \begin{tabular}{lrr}
\hline
\hline
BJD & $\rm RV_{\rm p}$ (km/s) & $\rm RV_{\rm s}$ (km/s) \\
\hline
56454.7903899  &    -55.59(49)   &  65.97(240) \\
56477.7589405  &    -62.57(20)   &  120.02(58) \\
56480.7366351  &    -61.58(73)   &  104.44(95) \\
56516.6366076  &    -57.78(119)  &   85.36(109)\\
56520.6163318  &    -34.32(50)   & -148.52(133)\\
56517.6346380  &    -28.95(115)  & -191.17(125)\\
56522.6128016  &    -61.52(15)   &  118.46(157)\\
56524.5905161  &    -29.46(27)   & -171.73(140)\\
56525.5797659  &    -61.14(168)  &  102.30(122)\\ 
56533.5901729  &    -30.73(32)   & -168.39(92) \\
56540.5215948  &    -29.07(55)   & -184.23(128)\\
\hline
    \end{tabular} 
\label{tab:rv}
\end{table}

%%%%%%%%%%%%%%%%%%%%%%%%%%%%%%%%%%%%%%%%%%%%%%%%%%

% Don't change these lines
\bsp	% typesetting comment
\label{lastpage}
\end{document}